\documentclass[preprint,3p,twocolumn]{elsarticle}
\usepackage{lineno,hyperref}
\modulolinenumbers[5]
\usepackage{graphicx}
\usepackage{paralist}
\usepackage{multirow}
\usepackage{amsmath} %
\usepackage{amssymb} 
\usepackage{subfigure}
\usepackage{wrapfig}
\usepackage{pgf,tikz,pgfplots}
\usepackage{standalone} 
\usepackage{hyperref}
\usepackage{textcomp}
\usepackage{xcolor}
\usepackage{siunitx}
\DeclareSIUnit\parsec{pc}
\DeclareSIUnit\lightyear{ly}
\DeclareSIUnit\HOUR{hour}
\DeclareSIUnit\DAY{day}
\DeclareSIUnit\YEAR{year}
\DeclareSIUnit\microBqperkg{\si[per-mode=symbol]{\micro\becquerel\per\kilo\gram}}
\DeclareSIUnit\milliBqperkg{\si[per-mode=symbol]{\milli\becquerel\per\kilo\gram}}
\DeclareSIUnit\picogperg{\si[per-mode=symbol]{\pico\gram\per\gram}}

\journal{Nucl.\ Instrum.\ Methods Phys.\ Res.\ A}

\begin{document}

\begin{frontmatter}

\title{Copper electroplating for background suppression\\in the NEWS-G experiment}

\author[queensUCanada]{L. Balogh}
\author[lpscGrenobleFrance]{C. Beaufort}
\author[queensUCanada]{A. Brossard}
\author[pnnlUSA]{R. Bunker}
\author[queensUCanada]{J.-F. Caron}
\author[queensUCanada]{M. Chapellier}
\author[queensUCanada]{J.-M. Coquillat}
\author[rmcCanada]{E.C. Corcoran}
\author[queensUCanada]{S. Crawford}
\author[lpscGrenobleFrance]{A. Dastgheibi Fard}
\author[uoaCanada]{Y. Deng}
\author[queensUCanada]{K. Dering}
\author[uoaCanada]{D. Durnford}
\author[queensUCanada]{G. Gerbier}
\author[ceaSaclayFrance]{I. Giomataris}
\author[queensUCanada]{G. Giroux}
\author[laurentianSudburyCanada,snolabCanada,McDonaldCanada]{P. Gorel}
\author[ceaSaclayFrance]{M. Gros}
\author[queensUCanada]{P. Gros}
\author[lpscGrenobleFrance]{O. Guillaudin}
\author[pnnlUSA]{E.W. Hoppe}
\author[uobUK]{I. Katsioulas}
\author[rmcCanada]{F. Kelly}
\author[ceaSaclayFrance,uobUK]{P.~Knights\corref{mycorrespondingauthor}}
\cortext[mycorrespondingauthor]{Corresponding author}
\ead{prk313@bham.ac.uk}
\author[rmcCanada]{L. Kwon}
\author[snolabCanada]{S. Langrock}
\author[subatechNantesFrance]{P. Lautridou}
\author[queensUCanada]{R.D. Martin}
\author[ceaSaclayFrance]{J.-P. Mols}
\author[lpscGrenobleFrance]{J.-F. Muraz}
\author[ceaSaclayFrance]{X.-F. Navick}
\author[uobUK]{T. Neep}
\author[uobUK]{K. Nikolopoulos}
\author[uoaCanada]{P. O'Brien}
\author[uobUK]{R. Owen}
\author[uoaCanada]{M.-C. Piro}
\author[lpscGrenobleFrance]{D. Santos}
\author[queensUCanada]{G. Savvidis}
\author[auotGreece]{I. Savvidis}
\author[queensUCanada]{F. Vazquez {de Sola Fernandez}}
\author[queensUCanada]{M. Vidal}
\author[uobUK]{R. Ward}
\author[lpscGrenobleFrance]{M. Zampaolo}

\address{(NEWS-G Collaboration)}

\author[pnnlUSA]{\normalsize\rm S.~Alcantar~Anguiano}
\author[pnnlUSA]{\normalsize\rm I.~J.~Arnquist}
\author[pnnlUSA]{\normalsize\rm M.L.~di Vacri}
\author[pnnlUSA]{\normalsize\rm K.~Harouaka}
\author[ICRR,IPMU]{\normalsize\rm K.~Kobayashi\fnref{wasedanow}}
\fntext[wasedanow]{Now at Waseda Research Institute for Science and Engineering, Waseda University, 3-4-1 Okubo, Shinjuku, Tokyo 169-8555, Japan}
\author[pnnlUSA]{\normalsize\rm K.S.~Thommasson}

\address[queensUCanada]{Department of Physics, Engineering Physics \& Astronomy, Queen's University, Kingston, Ontario K7L 3N6, Canada}
\address[lpscGrenobleFrance]{LPSC, Universit\'{e} Grenoble-Alpes, CNRS/IN2P3, Grenoble, France}
\address[pnnlUSA]{Pacific Northwest National Laboratory, Richland, Washington 99352, USA}
\address[rmcCanada]{Chemistry \& Chemical Engineering Department, Royal Military College of Canada, Kingston, Ontario K7K 7B4, Canada}
\address[uoaCanada]{Department of Physics, University of Alberta, Edmonton, Alberta, T6G 2R3, Canada}
\address[ceaSaclayFrance]{IRFU, CEA, Universit\'{e} Paris-Saclay, F-91191 Gif-sur-Yvette, France}
\address[laurentianSudburyCanada]{Department of Physics and Astronomy, Laurentian University, Sudbury, Ontario, P3E 2C6, Canada}
\address[snolabCanada]{SNOLAB, Lively, Ontario, P3Y 1N2, Canada}
\address[McDonaldCanada]{Arthur B. McDonald Canadian Astroparticle Physics Research Institute, Queen's University, Kingston, ON, K7L 3N6, Canada}
\address[uobUK]{School of Physics and Astronomy, University of Birmingham, Birmingham B15 2TT United Kingdom}

\address[subatechNantesFrance]{SUBATECH, IMT-Atlantique, Universit\'{e} de
  Nantes/IN2P3-CNRS, Nantes, France}
\address[auotGreece]{Aristotle University of Thessaloniki, Thessaloniki, Greece}
\address[ICRR]{Kamioka Observatory, ICRR, University of Tokyo,
  Higashi-Mozumi, Kamioka, Hida, Gifu 506-1205, Japan}

\address[IPMU]{Kavli Institute for the Physics and Mathematics of the
  Universe, University of Tokyo, Kashiwa, Chiba 277-8582, Japan}


\begin{abstract}
  New Experiments with Spheres-Gas (NEWS-G) is a dark matter direct
  detection experiment that will operate at SNOLAB (Canada). Similar
  to other rare-event searches, the materials used in the detector
  construction are subject to stringent radiopurity requirements. The
  detector features a 140-cm diameter proportional counter comprising two hemispheres made from commercially sourced 99.99\% pure
  copper.  Such copper is widely used in rare-event searches because
  it is readily available, there are no long-lived Cu radioisotopes,
  and levels of non-Cu radiocontaminants are generally low.
  However, measurements performed with a dedicated $^{210}$Po alpha
  counting method using an XIA detector 
  confirmed a problematic concentration of $^{210}$Pb in bulk of the
  copper.
  To shield the proportional counter's active volume, a
  low-background electroforming method was adapted to the
  hemispherical shape to grow a 500-\si{\micro}m thick layer of
  ultra-radiopure copper to the detector's inner surface. In this
  paper the process is described, which was prototyped at Pacific
  Northwest National Laboratory (PNNL), USA, and then conducted at
  full scale in the Laboratoire Souterrain de Modane in France. The
  radiopurity of the electroplated copper was assessed through
  Inductively Coupled Plasma Mass Spectrometry (ICP-MS). Measurements
  of samples from the first (second) hemisphere give $68\%$ confidence
  upper limits of $\textless0.58\;\si{\microBqperkg}$
  ($\textless0.24\;\si{\microBqperkg}$) and
  $\textless0.26\;\si{\microBqperkg}$
  ($\textless0.11\;\si{\microBqperkg}$) on the $^{232}$Th and
  $^{238}$U contamination levels, respectively. These results are
  comparable to previously reported measurements of electroformed
  copper produced for other rare-event searches, which were also found
  to have low concentration of $^{210}$Pb consistent with the background goals of
  the NEWS-G experiment.
\end{abstract}

\end{frontmatter}

\newpage

\section{Introduction}
Direct searches for dark matter (DM) and neutrinoless double-beta decay~\cite{Agnese:2016cpb,gerdaNature5442017, PhysRevD.98.102006,PhysRevLett.120.132501}
have strict requirements on the experimental background to achieve their targeted sensitivities.
While such experiments are generally carried out in underground
laboratories and in specifically designed shielding to suppress
backgrounds from external sources, one of the main remaining sources
arises from radioactive decays in the detector's construction
materials, including the gaseous target. 
The effort to procure materials with the lowest possible radioactivity has driven significant improvements in the techniques and facilities used to assay and prepare radiopure materials~\cite{SCOVELL2018160, LAFERRIERE201593, Warburton2004,Bunker:2020sxw}.

\begin{figure}[t!]
\centering
\includegraphics[width=0.9\linewidth]{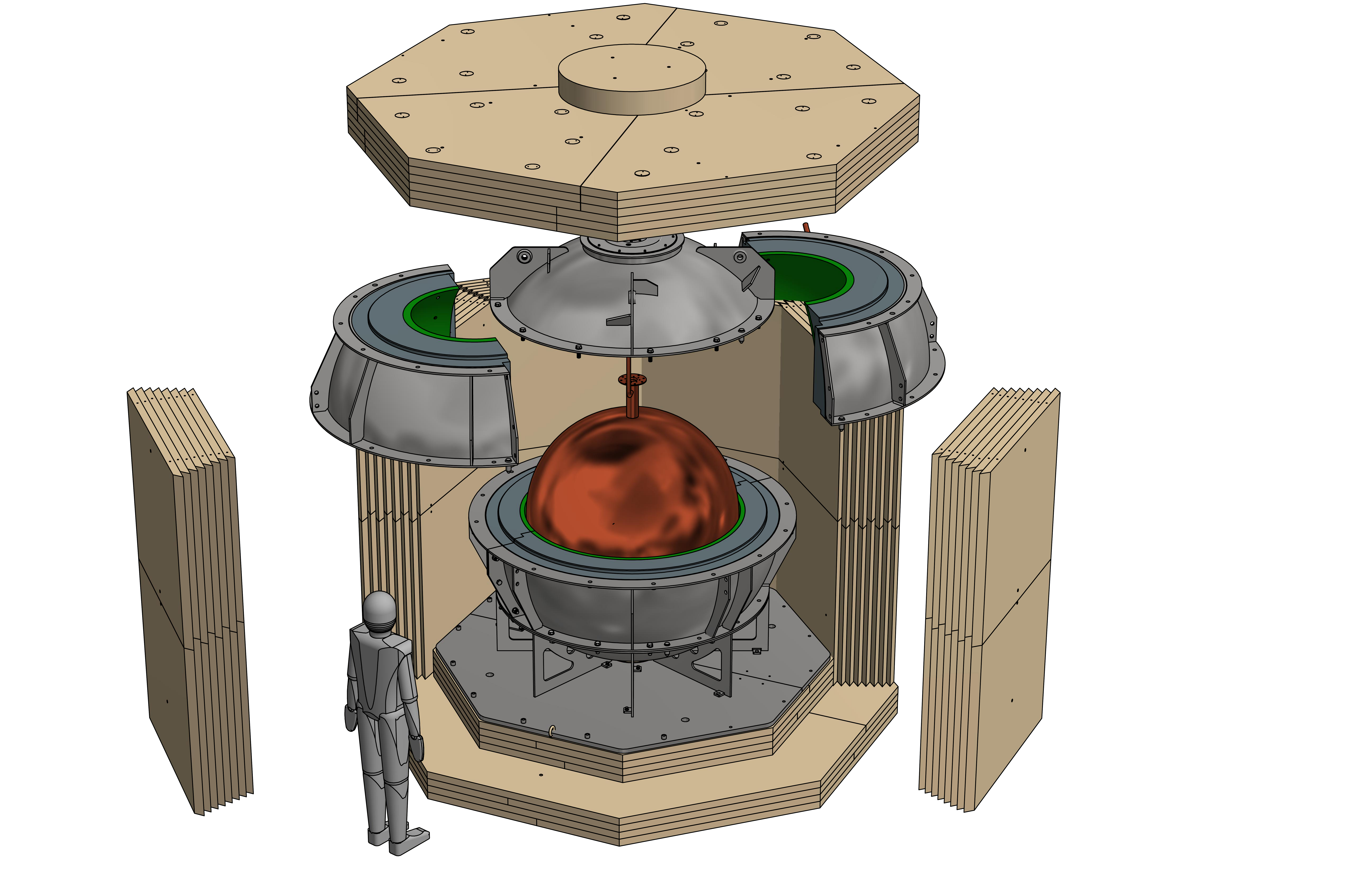}
\caption{Schematic diagram of the NEWS-G detector and shielding. The $\varnothing 140\;\si{cm}$ spherical
  proportional counter is shown at the centre, surrounded by 
  $3\;\si{cm}$ of archaeological lead, followed by $22\;\si{cm}$ of low
  radioactivity lead in a stainless steel skin. The
  outer-most part of the shielding comprises $40\;\si{cm}$ of
  high-density polyethylene (HDPE).\label{fig:newsgSNO}}
\end{figure}

A common choice for a high-purity material is commercially sourced copper~\cite{BUCCI2004132,ARMENGAUD2010294,Aalseth_2018}, 
because it is readily available and there are no long-lived Cu radioisotopes---with a half-life of $61.8\;\si{hours}$~\cite{nndcChartNuclides}, $^{67}$Cu is the longest-lived. 
For this reason, the NEWS-G collaboration~\cite{Arnaud2018-nr,Arnaud:2019nyp} chose 
C10100 ($99.99\%$ pure) copper\footnote{Procured from Aurubis AG, Hovestrasse 50, 20539 Hamburg, Germany} to construct a $\varnothing 140\;\si{cm}$ spherical proportional counter~\cite{Giomataris:2008ap,Savvidis:2016wei,Giganon:2017isb,Katsioulas:2018pyh,Giomataris:2020rna}, which will be housed in the compact shielding shown in Fig.~\ref{fig:newsgSNO}, to perform a direct DM search at SNOLAB, Canada. Along with this outer spherical shell, which is grounded, the detector is composed of a central electrode set at high voltage. To first approximation, this produces a radial 1/r$^{2}$ electric field in the detector. The interactions of a particle with the gas, such as a DM particle elastically interacting and resulting in a nuclear recoil, may cause ionisation. The resulting primary electrons drifting under the electric field until within approximately $1\;\si{\milli\meter}$ of the central electrode, where the electric field becomes large enough that the primary electrons gain sufficient energy to cause secondary ionisation. This results in a Townsend avalanche, thus providing signal amplification. Due to its proximity to the active medium and it's size, the radiopurity of the spherical shell is of critical importance. 

\begin{figure}[!t]
\centering
\includegraphics[width=0.85\linewidth]{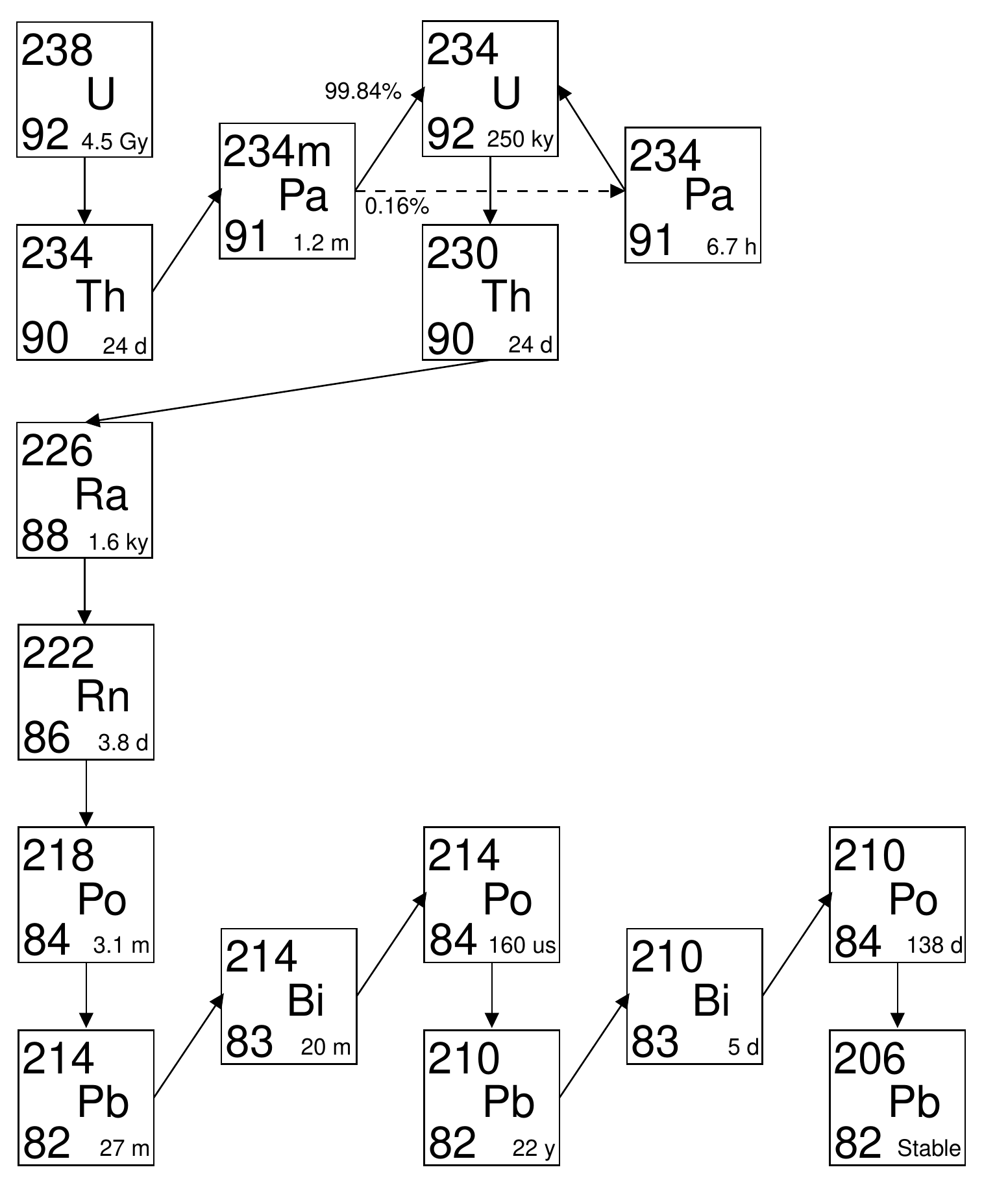}
\caption{${}^{238}$U decay chain. All daughters are solid at room temperature and pressure except $^{222}$Rn, which is a gas. Only decays with a branching fraction greater than $0.05\%$ are shown~\cite{nndcChartNuclides}.
\label{fig:u238DecayChain} }
\end{figure}

Even without long-lived Cu radioisotopes, a copper sample will have
some (non-Cu) radiogenic contamination resulting from cosmogenic
activation and industrial production processes.  For example,
cosmic-ray neutrons interacting with copper through the
($n$,\,$\alpha$) reaction can produce $^{60}$Co. The half-life of the
produced $^{60}$Co is approximately 5.3$\;\si{years}$, making it a long-lived background
relative to the typical time scale of direct DM detection
experiments. At the surface of the Earth, the added activity due to $^{60}$Co is approximately $0.4\;\si{\micro\becquerel\per\kilo\gram\per day}$~\cite{AlexisBrossard}. Other cosmogenic contaminants with shorter half-lives are
also produced, e.g.\ $^{59}$Fe. These contributions can be suppressed by minimising the copper's exposure to cosmic rays. Other radiocontaminants primarily originate from the $^{238}$U and $^{232}$Th decay chains.  The $^{238}$U decay chain is shown in Fig.~\ref{fig:u238DecayChain}. This contamination 
 is inherent to the raw material and a result of the manufacturing and handling processes. 
An established technique is to directly measure the uranium and thorium levels with inductively coupled plasma mass spectrometry (ICP-MS), which has been demonstrated 
to have sensitivity better than $30\;\si[per-mode=symbol]{\femto\gram\per\gram}$ to these contaminants~\cite{LAFERRIERE201593,ARNQUIST2020163761,ABGRALL201622}.  
The progeny activities can also be inferred and used to estimate background contributions to experiments, under the assumption of secular equilibrium.

However, $^{222}$Rn, which is part of the $^{238}$U decay chain, is a
gaseous isotope. As a result, $^{222}$Rn may deposit its decay
products on the copper surface or into the copper bulk at the raw-ore
stage or during manufacturing. This contribution adds to the
contamination and may break the secular-equilibrium assumption.
The longest-lived isotope in the $^{222}$Rn decay chain is $^{210}$Pb
with a half-life of
$22.2\;\si{years}$~\cite{SHAMSUZZOHABASUNIA2014561}. Accumulation of
$^{210}$Pb from ${}^{222}$Rn deposits can result in experiment
backgrounds that cannot be inferred by ICP-MS measurements of the
$^{238}$U progenitor.
One method to assess this contamination is by directly measuring the $5.3\;\si{\mega\eV}$ $\alpha$ particles from the $^{210}$Po decays~\cite{Bunker:2020sxw,ZUZEL2017165,ABE2018157}, using a high-sensitivity XIA UltraLo-1800 spectrometer, which has a sensitivity of $0.0001\;\alpha\si{\per {\centi\meter}^{2} \per hour}$~\cite{xiacounter}. 
The XMASS collaboration has established a method to estimate very low  $^{210}$Pb contamination in copper bulk, having demonstrated the ability to distinguishing the contamination in bulk from that on the surface~\cite{ABE2018157}.
For oxygen-free copper (at least $99.96\%$ pure by weight\footnote{Japanese Industrial Standard, JIS:C1020}) the $^{210}$Pb contamination in the bulk is estimated to be in the range of 17--40$\;\si{\milliBqperkg}$~\cite{ABE2018157}. The corresponding value for the C10100 copper procured by the NEWS-G collaboration is $29^{+8+9}_{-8-3}\;\si{\milli\becquerel\per\kilo\gram}$, as discussed in Section~\ref{sec:Pb210InOurCopper}.

The measured level of ${}^{210}$Pb in the C10100 copper bulk of the NEWS-G detector and the corresponding contamination of its progeny would represent approximately $82\%$ of the experimental background~\cite{AlexisBrossard} below $1\;\si{\kilo\eV}$ as estimated by means of a Geant4~\cite{Agostinelli:2002hh} simulation, excluding possible contributions originating from activation of the copper induced by the surface-level flux of cosmic muons. 
An approach to suppress the background from $^{210}$Pb contamination
is to grow a layer of ultra-radiopure copper onto the inner surface
of the detector sphere. This layer acts as an internal shield to
suppress backgrounds, e.g.\ from $\beta$-decays of $^{210}$Pb and
accompanying X-rays and Auger electrons, and its
progeny $^{210}$Bi, originating from the bulk of the commercially
sourced C10100 copper.
It is estimated that a
$500\;\si{\micro\meter}$-thick layer of ultra-radiopure copper will
suppress this background contribution below $1\;\si{\kilo\eV}$ by
a factor of 2.6~\cite{AlexisBrossard}.

A method to deposit ultra-radiopure copper is potentiostatic electroforming~\cite{ABE2018157, Hoppe:2014nva,Hoppe2008}. This method takes advantage of electrochemical properties to produce copper with reduced impurities. The process is described in Section~\ref{sec:electroplating}. This method was previously used to produce a variety of detector components, including those requiring extreme radiopurity such as for the Majorana Demonstrator~\cite{Abgrall:2013rze}. 
Internal fittings were fabricated from electroformed copper with $^{238}$U and $^{232}$Th levels less than   
0.099 and 0.119$\;\si{\microBqperkg}$\footnote{For ${}^{238}$U, $1\;\si[per-mode=symbol]{\micro\becquerel \per\kilo\gram} \approx 0.081\;\si{\picogperg}$.  
For ${}^{232}$Th, $1\;\si{\microBqperkg} \approx 0.244\;\si{\picogperg}$} at 68\% confidence, respectively---limited by the ICP-MS assay precision~\cite{ABGRALL201622}. The $^{210}$Pb contamination of electroformed copper has previously been measured to be below the sensitivity of an XIA UltraLo-1800 spectrometer, with a $90\%$ confidence level upper limit on its activity of ${<}5.3\;{ {\rm mBq/kg}}$~\cite{ABE2018157}. In order to apply this process to a hemispherical surface, a scale model was produced and used to determine the operating conditions. This is described in Section~\ref{sec:scaleModel}. The electroplating procedure used on the NEWS-G detector and the results of a subsequent radioisotope assay of the produced copper are detailed in Section~\ref{sec:electroplatingNEWSGDetector} and Section~\ref{sec:assayResults}, respectively. 

\section{Assessment of the $^{210}$Pb Contamination in NEWS-G Copper}
\label{sec:Pb210InOurCopper}
To assess the level of $^{210}$Pb contamination in the C10100 copper used to produce the detector, samples were taken from the same batch of copper after casting. %
The $\alpha$ particles from $^{210}$Po decays were measured using an XIA UltraLo-1800~\cite{xiacounter} ionisation chamber, which uses an active veto to obtain a second complementary signal arising from cases where the $\alpha$ particle does not originate from the sample under test. This is used to suppress background coming from the detector's own construction materials. The sample is placed in the detector which is flushed with argon gas to minimise $^{222}$Rn contamination. 
In this measurement, the $^{210}$Po content of the bulk of the copper sample
is of interest. The observable energy of $5.30\;\si{\mega\eV}$ $\alpha$ particles emerging from the bulk of the copper sample was estimated with a Geant4 simulation. An energy window of $2.5\;\si{\mega\eV}$ to $4.8\;\si{\mega\eV}$ was used to primarily select $\alpha$ particle originating from a depth of approximately $2\;\si{\micro\meter}$ to $8\;\si{\micro\meter}$. 
This improves the signal-to-noise ratio for selecting bulk $\alpha$ particle events. 
The conversion factor for measured counts to bulk activity was estimated from Geant4 to be $\num{2.7E2}\;\si{(\becquerel\per\kilo\gram)\per(\alpha{\per {\centi\meter}^{2} \per hour)}}$~\cite{ABE2018157}.

$^{210}$Po has a half-life of approximately $138\;\si{days}$, which is significant shorter than the approximately $22\;\si{years}$ of the progenitor $^{210}$Pb. As a result, the activities of $^{210}$Po and $^{210}$Pb may be different due to different contamination amounts at the production phase. Therefore, the activities of the two isotopes may be out of secular equilibrium; however, the $^{210}$Po activity in a sample will evolve over time until it matches that of $^{210}$Pb. Therefore, multiple measurements of the $^{210}$Po activity over time are required 
to accurately infer the activity of $^{210}$Pb in the copper. Four measurements of the $\alpha$ particles from the sample were made over the course of approximately one year, each lasting between 12 and 23 days. Table~\ref{tab:xiaMeasurements} shows the results of the four measurements. 

\begin{table}[!h]
\caption{Measurements of the $\alpha$ particles in a $2.5\;\si{\mega\eV}$ to $4.8\;\si{\mega\eV}$ energy window originating from ${}^{210}$Po decays in a C10100 copper sample.}
\label{tab:xiaMeasurements}
\begin{tabular}{lc}
\hline
Date                         & Measurement \\
 & {[}\num{E-4}~$\alpha\si{\per {\centi\meter}^{2} \per hour}${]} \\ \hline
Jul. 2 - 25, 2018            & $2.3\pm0.4$                           \\
Oct. 5 - 17, 2018            & $2.2\pm0.4$                             \\
Dec. 28, 2018 - Jan. 9, 2019 & $1.4\pm0.3$                             \\
Apr. 19 - May 7, 2019        & $1.4\pm0.3$                             \\ \hline
\end{tabular}
\end{table}

\begin{figure}[!h]
\centering
\includegraphics[width=0.85\linewidth]{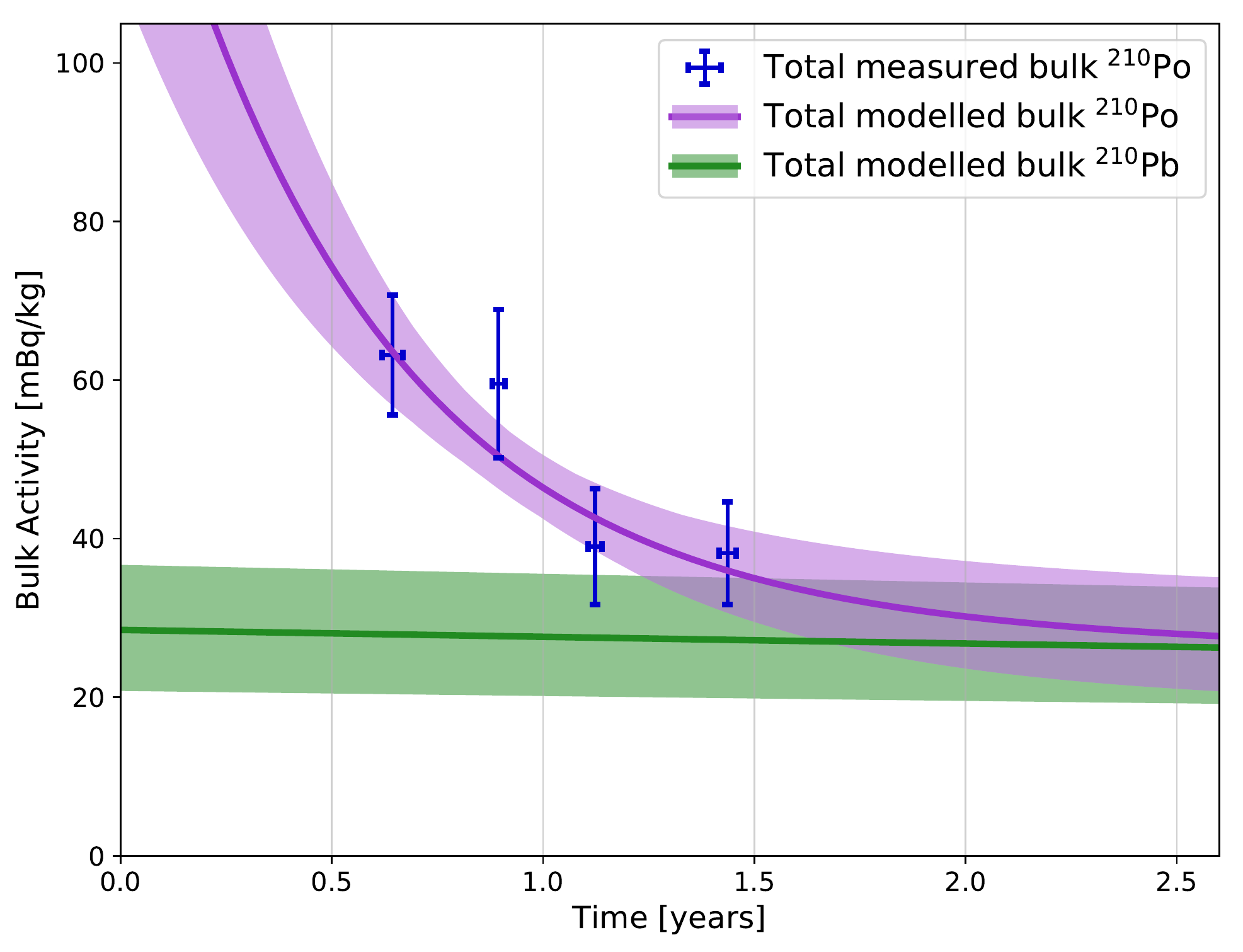}
\caption{Measurements of the $\alpha$ particles from the decay of ${}^{210}$Po in a sample of C10100 copper used in the production of the NEWS-G detector. Time is measured from the estimated production date of the copper. The purple (green) line shows the fitted ${}^{210}$Po (${}^{210}$Pb) activity over time, with the bands showing the $\pm1\sigma$ region.
\label{fig:xiaMeasurementsFit} }
\end{figure}

 A joint likelihood fit of all measurements was performed and is shown
 in Figure~\ref{fig:xiaMeasurementsFit} along with the
 measurements. From this fit, it was estimated that the $^{210}$Pb
 activity in the sample is
$29^{+8+9}_{-8-3}\;\si{\milli\becquerel\per\kilo\gram}$,
 where the statistical and systematic uncertainties are given
 separately.
 This is consistent with other copper samples with similar purity~\cite{ABE2018157}.

\section{Electroplating}
\label{sec:electroplating}
\begin{figure}[!h]
\centering
\includegraphics[width=0.85\linewidth]{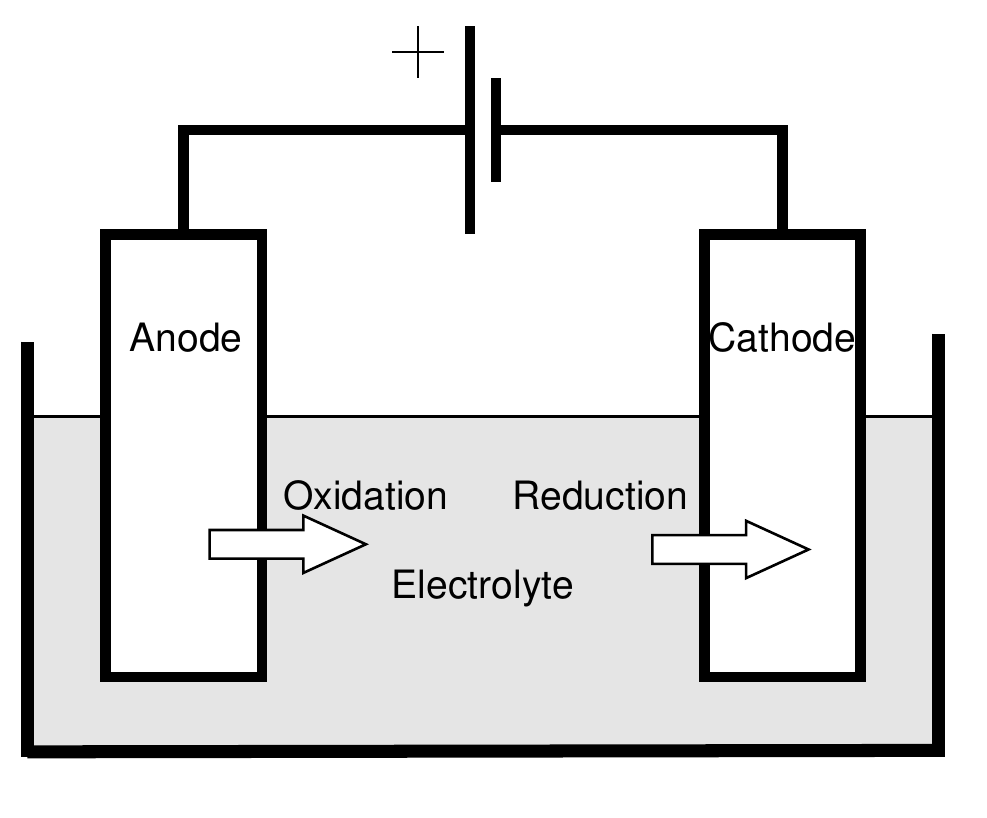}
\caption{Schematic diagram of a simple electrolytic cell. Arrows indicate the motion of ions, which are released into the electrolyte by oxidation reactions at the anode and then deposited on the cathode in reduction reactions. 
\label{fig:electrolyticCell}}
\end{figure}

Electroplating is carried out through the use of an electrolytic cell, which
consists of an anode and a cathode separated by an electrolyte, as illustrated in Fig.~\ref{fig:electrolyticCell}.
A current is used to supply electrons to the cathode where an ion undergoes a reduction reaction (gain of electrons) to form an atom deposited on the surface, while oxidation reactions (loss of electrons) occur at the anode. 
The reactions occurring at each of the electrodes will be of the general form:
\begin{equation}
A^{(y+z)+} + z e^{-} \leftrightharpoons A^{y+}\,,
\end{equation}
where $A$ is the molecular species, $y$ is its ionic charge and $z$ is the number of electrons required for the reduction reaction (reading left-to-right) or the number of electrons released in the oxidation reaction (reading right-to-left). Reading this equation in one direction gives the ``half-cell reaction'', where the anode and cathode half-cell reactions are not necessarily the same; e.g., in the case where one species is oxidised at the cathode but a different species is reduced at the anode. 

A current $I$ flows through the circuit and electrolyte. As the reduction reactions require electrons, the number of moles of reduced atoms during electroplating is proportional to the total supplied charge: $Q(t)=\int I \mathrm{d}t$.  
The number of moles $n$ of ions reduced in time $t$ is given by
\begin{equation}
n(t) = \frac{Q(t)}{zF} \,,
\label{eq:numberMolesDeposited}
\end{equation}
 where $F=eN_{A}$ is the Faraday constant, and $e$ and $N_{A}$ are the elementary charge and the Avogadro constant, respectively.
 The resulting deposited mass as a function of time is 
\begin{align}
M(t) &= m_{r} n(t)  \,,
\label{eq:massDeposited}
\end{align}
where $m_{r}$ is the molecular mass of the deposited species. 
When the current is reversed the process is called electropolishing, which is a technique used to remove material from a surface.

There will be several species of ions in the electrolyte available to electroplate to the cathode.
The tendency of an ion species to be reduced is quantified by the reduction potential $E^{0}$. Examples are shown in Table~\ref{tab:SEPtable} for copper and radioisotope contaminants. A greater value of $E^{0}$ indicates a species that is more easily reduced. 
Each half-cell reaction will have its own reduction potential. The standard cell potential $E^{0}_{\text{cell}}$ of the electrolytic cell is defined as the difference between the reduction potentials of the half-cell reactions at the anode $E_{\text{A}}^{0}$ and cathode $E_{\text{C}}^{0}$:
\begin{equation}
E^{0}_{\text{cell}} = E^{0}_{\text{C}} - E^{0}_{\text{A}}\,.
\end{equation}
For $E^{0}_{\text{cell}}<0$, additional energy will be required for the reaction to proceed~\cite{greiner1995thermodynamics}. For $E^{0}_{\text{cell}}\geq0$, the reaction is spontaneous (or in chemical equilibrium in the case of equality).
For a given species being oxidised at the anode, the reaction will only proceed when the cathode half-cell reaction has a higher reduction potential. In the case of a copper anode being oxidised, only ion species in the electrolyte with a reduction potential greater than that of copper will reduce. The relatively high reduction potential of copper compared to many radioisotopes means that it is purified during electroplating. However, other factors, such as mass transport of contaminant ions, can cause species with lower reduction potentials to be deposited with the copper in small amounts~\cite{Hoppe2008}.

\begin{table}[!h]
\centering
\caption{Reduction potential for copper and possible radiocontaminants. \label{tab:SEPtable}}
\vspace{0.5em}
\begin{tabular}{lllll}
\hline
Reductants &  & Oxidants & $E^{0}~(\si{\volt})$ & \\ \hline
       	$\text{Cu}^{2+} + 2 e^{-}$  & $\leftrightharpoons$  	&$\text{Cu}$   	& $ +0.34$&\cite{bard1985standard} \\
          $\text{Pb}^{2+}+2e^{-}$	&$\leftrightharpoons$	&$\text{Pb}$	& $-0.13$&\cite{atkins1997crc}     \\    
         $\text{U}^{3+}+3e^{-}$	&$\leftrightharpoons$	&$\text{U}$	& $-1.80$&\cite{lide2006crc}     \\     	
                   $\text{Th}^{4+}+4e^{-}$	&$\leftrightharpoons$	&$\text{Th}$	& $-1.90$&\cite{lide2006crc}    \\    	
         $\text{K}^{+}+e^{-}$    	&$\leftrightharpoons$	&$\text{K}	$       & $-2.93$&\cite{haynes2011crc}     \\   \hline  
       \end{tabular}
\end{table}

In this work, a copper anode is used to provide Cu$^{2+}$ ions to the electrolyte. For Cu${}^{2+}$ ions reducing at the cathode, the system will have $E^{0}_{\text{cell}}=0\;\si{\volt}$. Thus, to drive the reaction and overcome energy loss mechanisms in the system, the electrodes are kept at a potential difference of $0.3\;\si{\volt}$~\cite{Hoppe2009JRADIOANALNUCLCHEM}.

\begin{figure}[!h]
\centering
\includegraphics[width=0.95\linewidth]{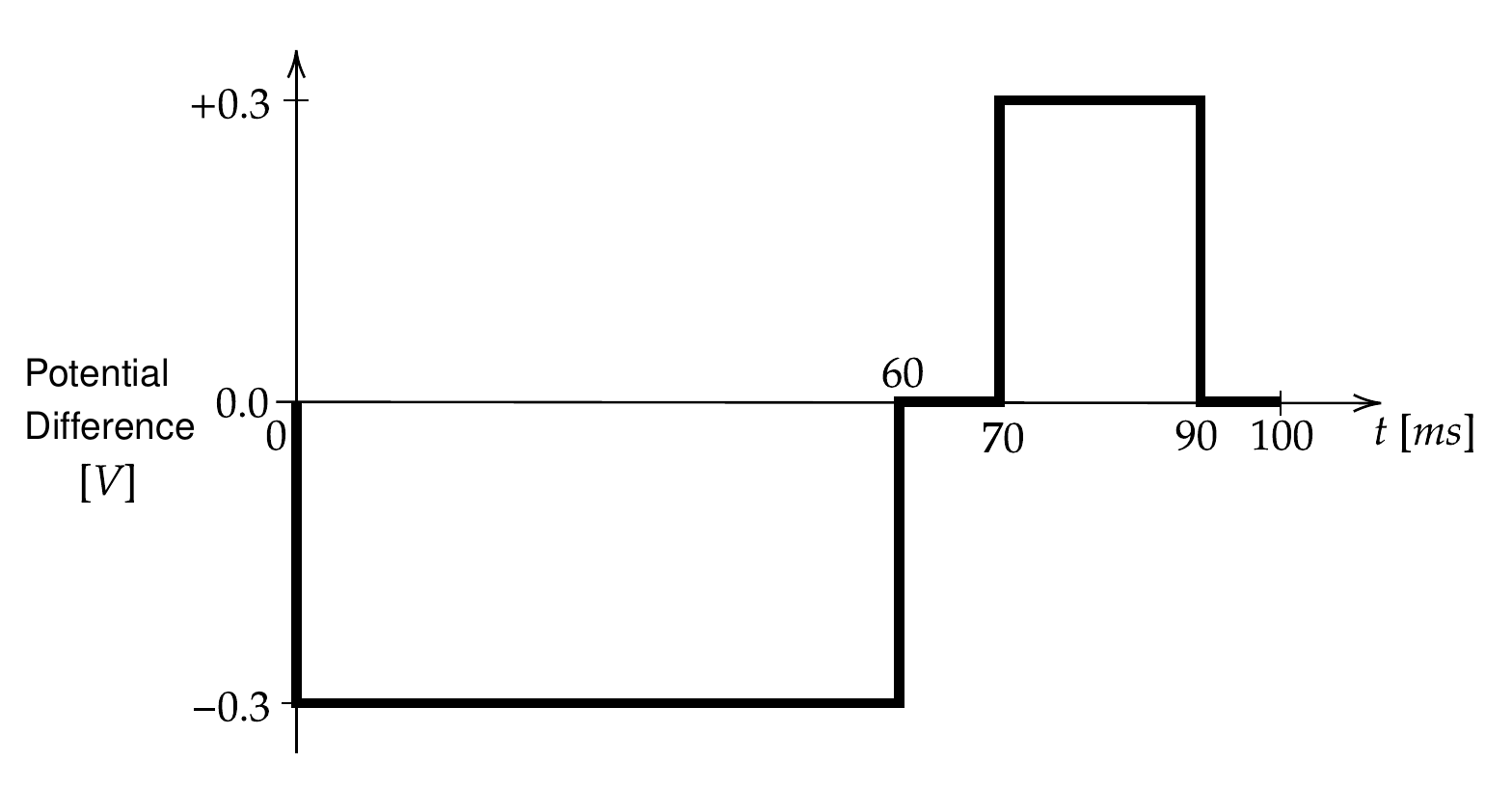}
\caption{Waveform used in the electroplating. The negative terminal was attached to the cathode. 
\label{fig:platingPulse} }
\end{figure}

It has been shown that applying a time-varying potential difference between the electrodes can have several benefits compared to a constant potential difference~\cite{CHANDRASEKAR20083313}.
During electroplating, the region of the electrolyte at the surface of the cathode becomes depleted of Cu$^{2+}$ relative to the bulk electrolyte.  
This slows down the rate of electroplating and affects the properties of the deposited copper~\cite{CHANDRASEKAR20083313}. The waveform of the time-varying potential difference allows this region to be replenished by allowing diffusion from the bulk electrolyte when no voltage is applied and by reintroducing more ions from the surface during the reverse-voltage part of the waveform. 
Also, differences in current density can arise due to differences in the distance between the anode and cathode surfaces (e.g.\ a surface rough point). High current density regions of the electrolyte are more depleted of Cu$^{2+}$ than lower density regions.  
When no voltage is applied, ions can diffuse between two such regions and thus lead to a more uniform overall current density, while the reverse-voltage part of the waveform prevents a thick layer forming in the high current density regions~\cite{CHANDRASEKAR20083313}; both effects promote more uniform growth of the electroplated copper layer. The reversing of polarity also allows for release of contaminant ions that may have been entrapped during the high mass transport portion of the forward plating.
The waveform used for the electroplating is shown in Fig,~\ref{fig:platingPulse}.
Note that while the potential is applied it is potentiostatic at a level that favours the oxidation/reduction of copper.

\section{Scale Model}
\label{sec:scaleModel}
The copper electroplating procedure described in the previous section
is a well-established and successful method that has existed for over
a decade~\cite{Hoppe2008}. However, this method must meet fairly rigid
operational conditions to produce optimal material. Failure to meet
these conditions can not only produce copper of poor radiopurity but
often results in deposits with poor physical properties as well.
For NEWS-G the initial loading of copper into solution would
need to be generated from an initial electropolishing step, because
commercially available copper sulphate is not sufficiently
pure. However, for traditional electroforming, the amount of copper
required in the electrolyte is too great to achieve through electropolishing.
As a result, the plating conditions for the NEWS-G hemispheres
required a major deviation in the concentration of copper sulphate
(CuSO$_{4}$) in the electrolyte.

Not all parameters have a well-studied effect on growth, especially
when multiple parameters are outside of their established optimal
operating ranges. Prior experience has shown that electrolyte with a
low copper-ion concentration can produce dendritic copper
deposition. In the absence of accurate deposition models it was
necessary to run a scaled experiment prior to plating the full-sized
$\varnothing 140\;\si{cm}$ sphere underground in Laboratoire
Souterrain de Modane (LSM). Key growth
parameters were identified and an experiment was designed based on
those that could be adjusted \textit{in situ} at LSM and projected
onto a scale model.

The key independent and adjustable variables were determined to be the
concentration of copper and overall conductivity of the electrolyte,
and the current based on the limiting set of voltage
conditions. Control of the CuSO$_{4}$ concentration is limited by the
amount of copper that can be dissolved during an initial
electropolishing step, which serves two purposes:
\begin{inparaenum}[a)]
\item expose the underlying bulk crystal structure to prepare the
  copper surface for electroplating; and
\item load the electrolyte with copper.
\end{inparaenum}
During this step, the $\varnothing 140\;\si{cm}$ hemisphere will act
as the anode and careful control of the potential is not as
important, whereas subsequently copper will be plated to the
$\varnothing 140\;\si{cm}$ hemisphere which will then be serving as
the cathode.
During the latter step, the voltage control and deposition rate are
critical. As a result, establishing how the plating responds to small changes in
CuSO$_{4}$ is crucial. As such, three
variations of CuSO$_{4}$ concentration, three conductivities, and
three voltage settings were identified for experimentation on the
scale model.

\begin{figure}[t!]
\centering
\subfigure[\label{subfig:ScaleModel}]{\includegraphics[width=0.45\linewidth]{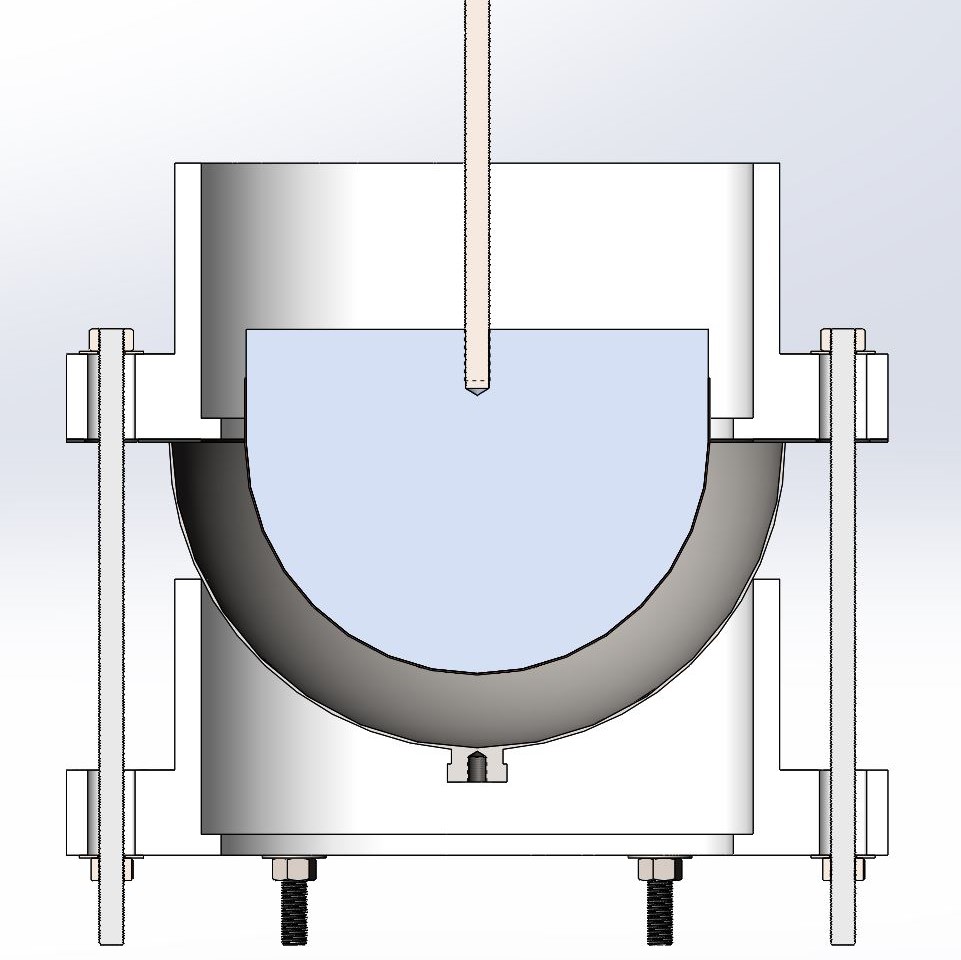}}
\subfigure[\label{subfig:BuiltScale}]{\includegraphics[width=0.45\linewidth]{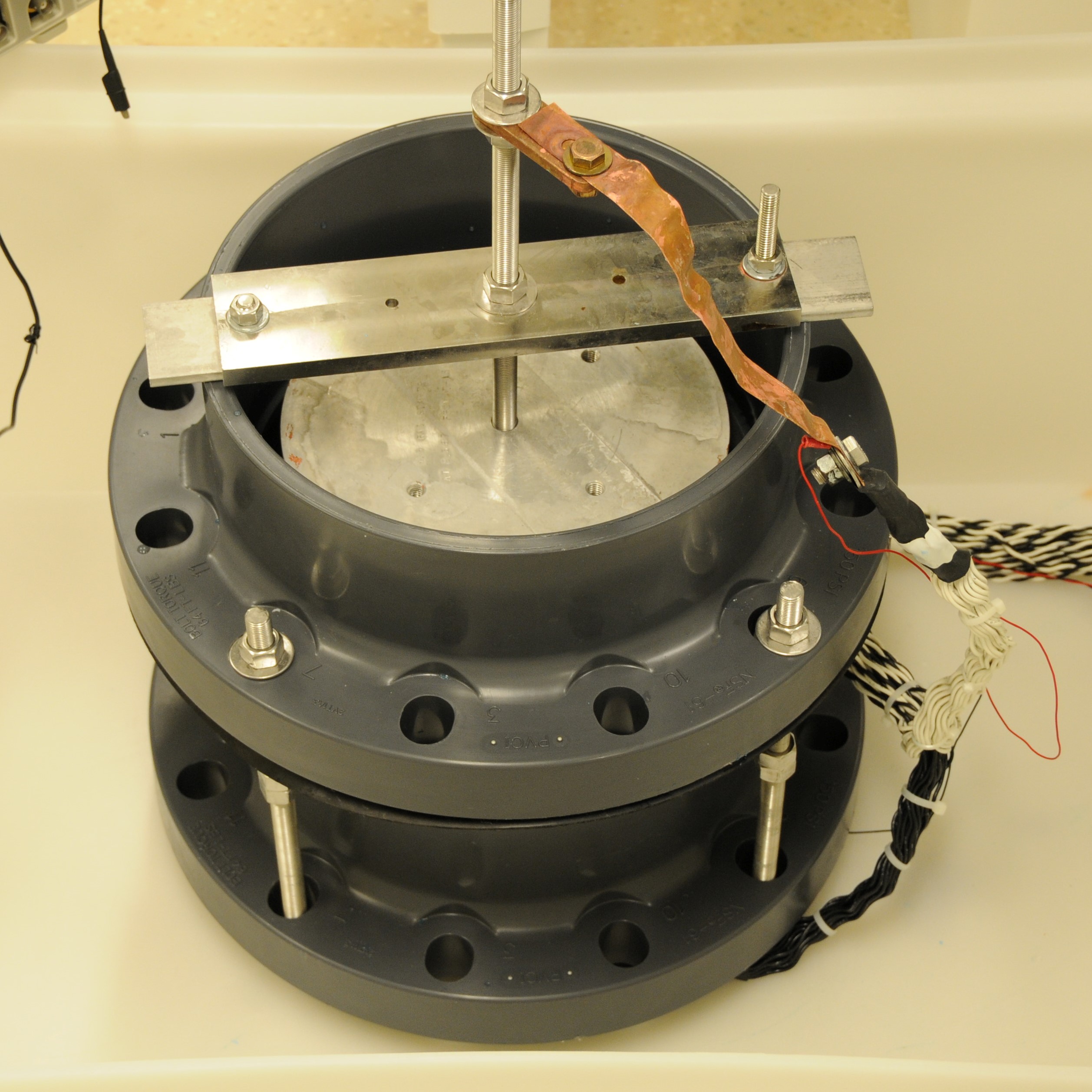}}
\subfigure[\label{subfig:GrowthScale}]{\includegraphics[width=0.45\linewidth]{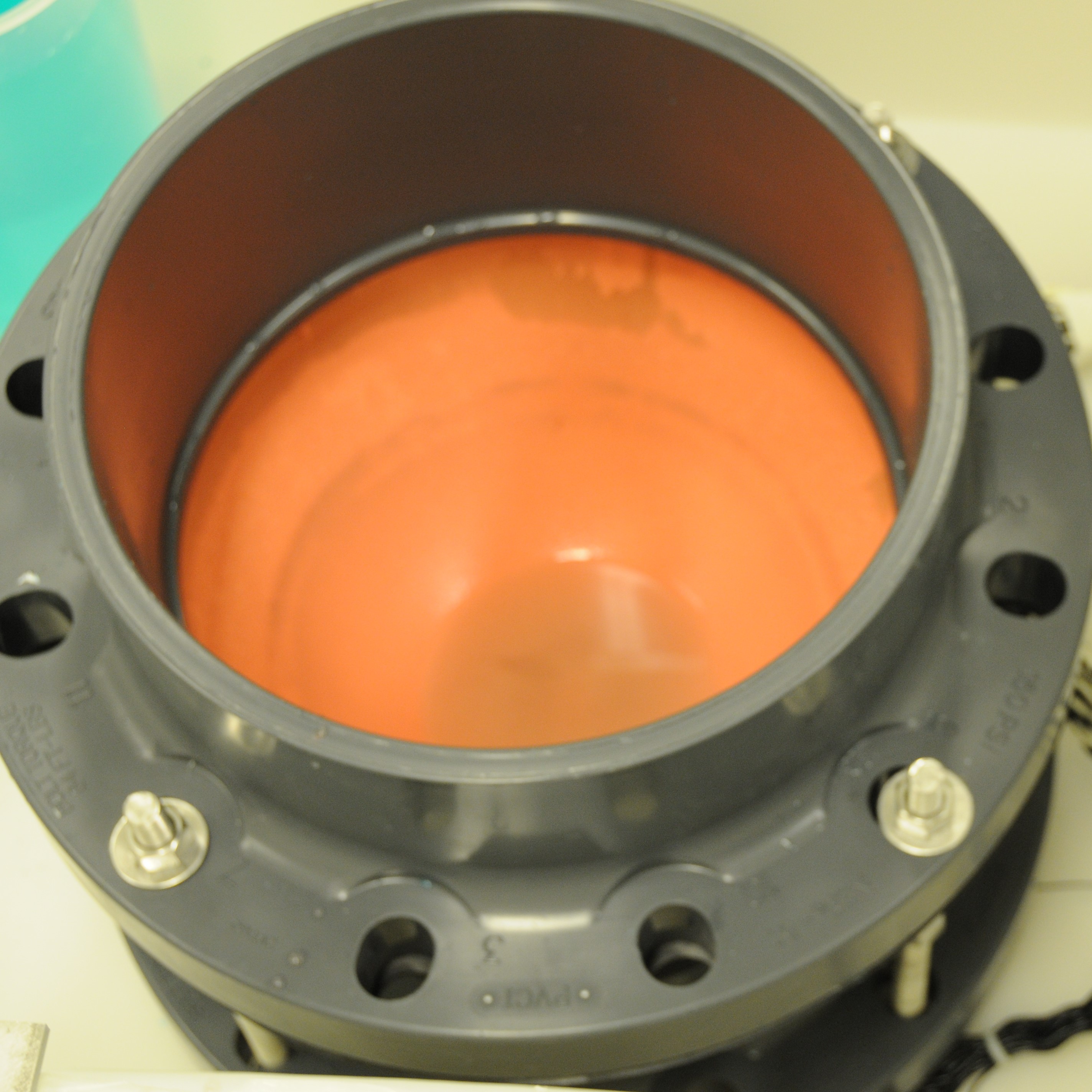}}
\subfigure[\label{subfig:FinalScale}]{\includegraphics[width=0.45\linewidth]{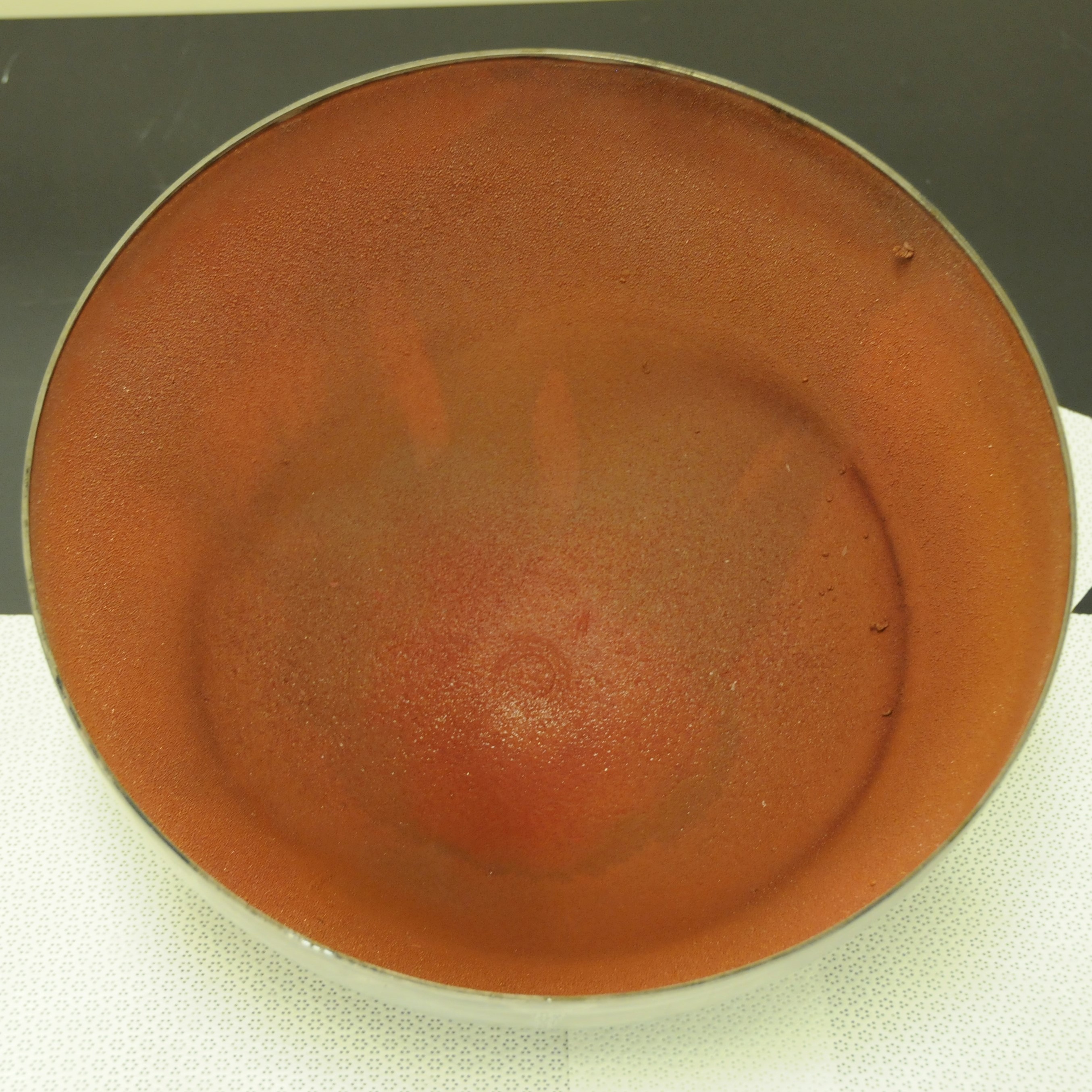}}
 \caption{\subref{subfig:ScaleModel}\,CAD model of the small-scale setup; \subref{subfig:BuiltScale}\,the assembled scale-model experiment; \subref{subfig:GrowthScale}\,copper plated onto the scale model's stainless steel hemisphere; and \subref{subfig:FinalScale}\,the final scale-model growth of copper.
 \label{fig:ScaleEx}}
\end{figure}

A stainless steel spherical float with a diameter of 30 cm was cut in
half and used as a stand-in for the full-scale
$\varnothing 140\;\si{cm}$ copper hemisphere. A smaller hemisphere was
machined from aluminium and plated with copper to serve as the anode
after the initial electropolishing step.  Figure~\ref{fig:ScaleEx} shows the
experimental setup of the scale model. Although the transport dynamics
involved are not fully understood, previous experience has shown that
the electrode gap (path length) has an effect on plating, regardless
of CuSO$_{4}$ concentration and conductivity. So, while the spacing
between the two electrodes was scaled, the impedance needed to be
matched to that of the full-scale setup. This required the electrolyte
conductivity to be reduced to compensate for the reduced electrode
spacing in the model.

A bath, shown in Fig.~\ref{subfig:ScaleModel} and Fig.~\ref{subfig:BuiltScale}, was designed to hold and stabilise the stainless steel hemisphere. Several iterations of plating were performed to cycle through the plating variations and determine the optimal electroplating conditions based primarily on the quality of grain uniformity and size observed in the deposit. Based on these trials, the parameters chosen for plating copper onto the full-scale hemispheres were a CuSO$_{4}$ concentration of $0.03\;\si{\mole\per\liter}$, a conductivity of $91.9\;\si{\milli\siemens\per\centi\meter}$ (corresponding to a full-scale conductivity of $300\;\si{\milli\siemens\per\centi\meter}$), and a potential of $0.35\;\si{\volt}$.  Using these parameters, the estimated time to electroplate each full-scale hemisphere was $\sim$8 days to attain a thickness of $500\;\si{\micro\meter}$. The resulting growth for the small-scale model is shown in Fig.~\ref{subfig:GrowthScale} and Fig.~\ref{subfig:FinalScale}. 

\section{Electroplating the NEWS-G Detector}
\label{sec:electroplatingNEWSGDetector}
\begin{figure}[t!]
\centering
\subfigure[\label{subfig:cleanedLQ}]{\includegraphics[width=0.49\linewidth]{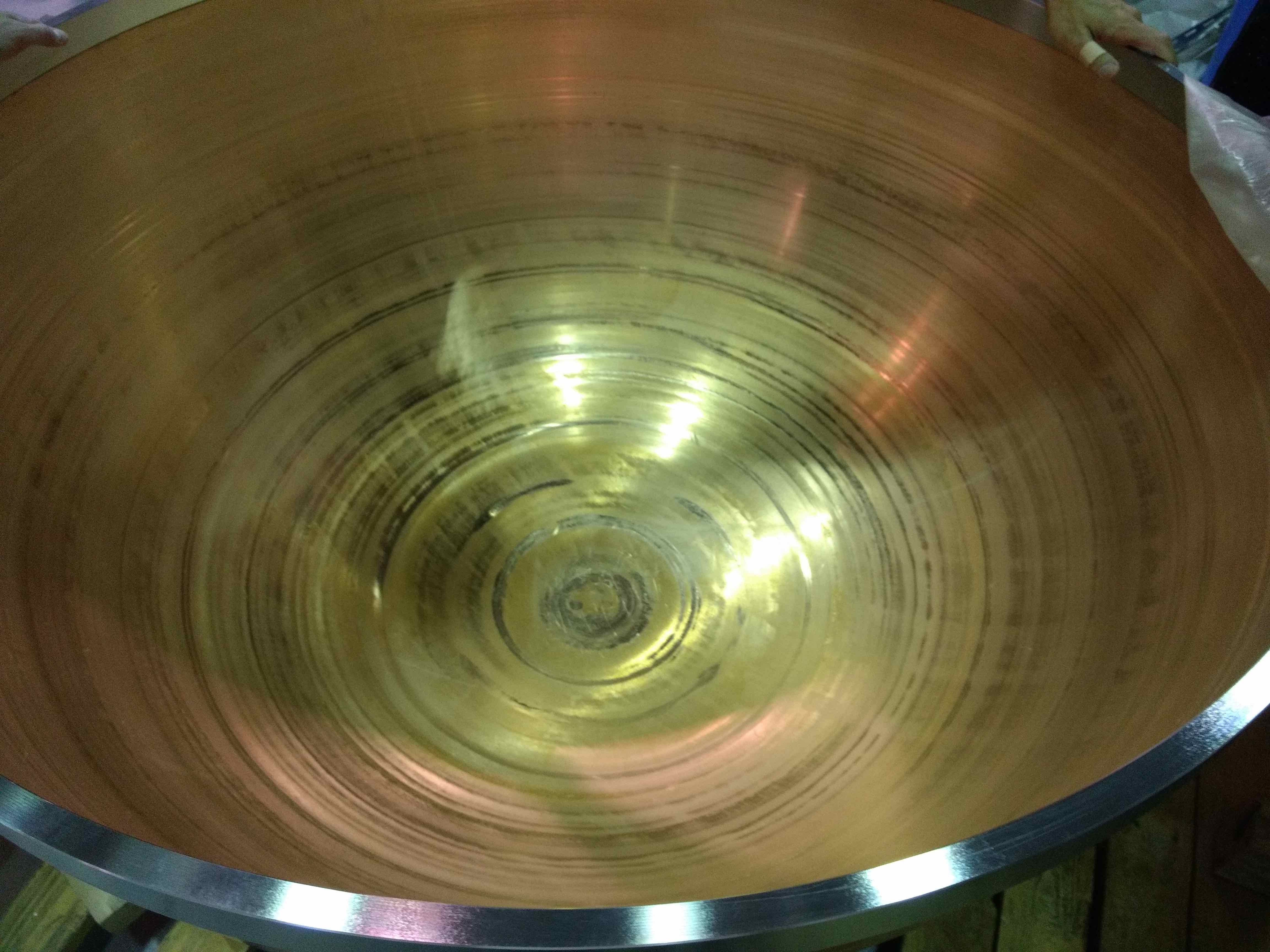}}
\subfigure[\label{subfig:etchedLQ}]{\includegraphics[width=0.49\linewidth]{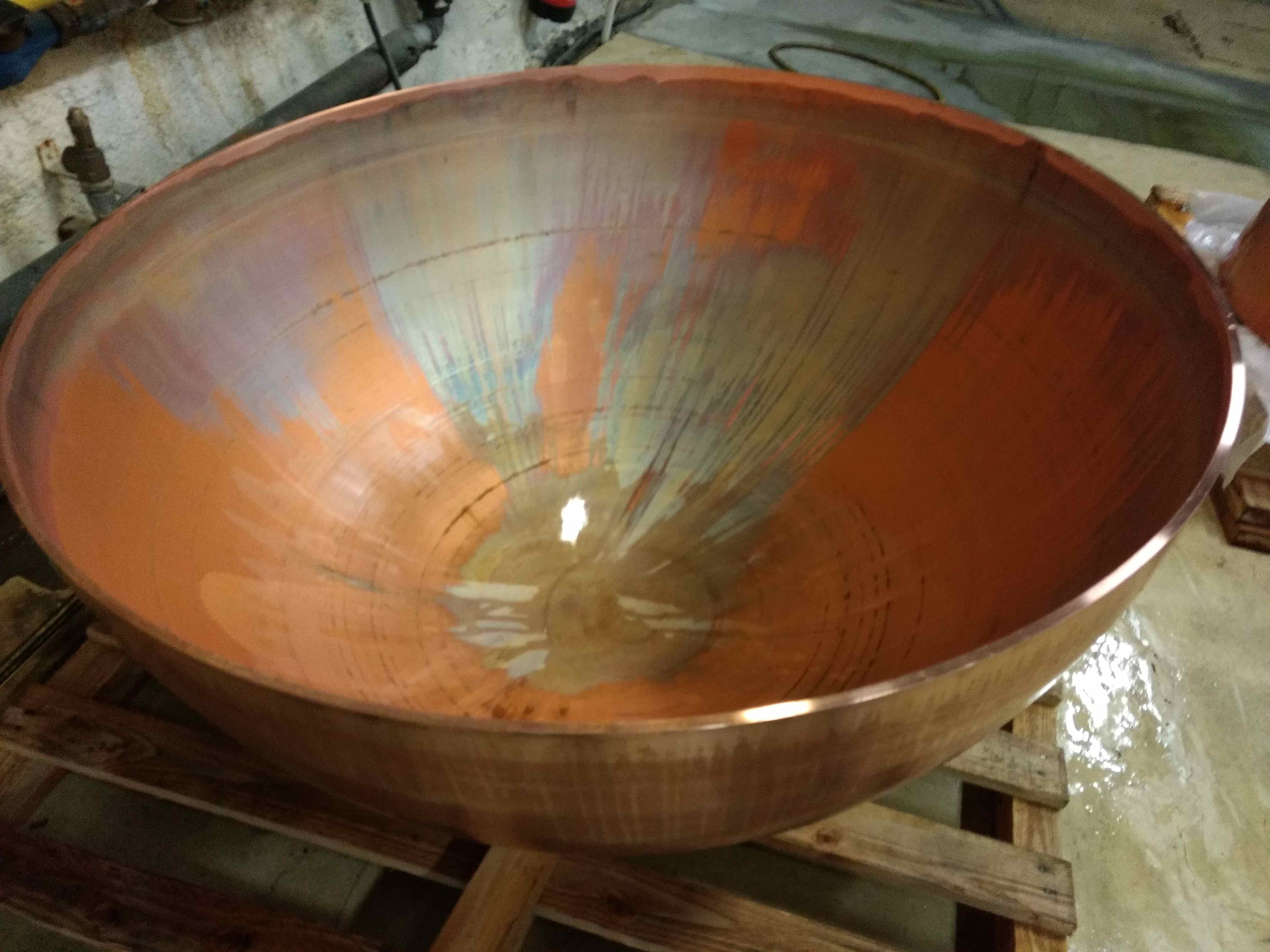}}
  \caption{A detector hemisphere following \subref{subfig:cleanedLQ}\,initial cleaning with detergent and \subref{subfig:etchedLQ}\,chemical etching with an acidified hydrogen peroxide solution. The discolouration observed in the latter is a result of oxidisation of the copper and it is removed when the hemisphere is put in contact with the electrolyte. 
 \label{fig:beforePlating}}
\end{figure}

The electroplating was conducted at LSM at a depth of
$4800\;\si{\meter}$ water equivalent to reduce cosmogenic activation. The detector outer shell is comprised of two $\varnothing 140\;\si{cm}$ hemispheres, produced by a spinning technique using C10100 copper. The result after cleaning with commercial detergent is shown in Figure~\ref{subfig:cleanedLQ}. The hemispheres were then sanded 
to produce a smooth surface and subsequently chemically etched using an acidified hydrogen peroxide solution~\cite{Bunker:2020sxw,HOPPE2007486}. The result of this preparation is shown in Figure~\ref{subfig:etchedLQ}. 

\begin{figure}[b!]
\centering
\subfigure[\label{subfig:hemi1Installed}]{\includegraphics[width=0.80\linewidth]{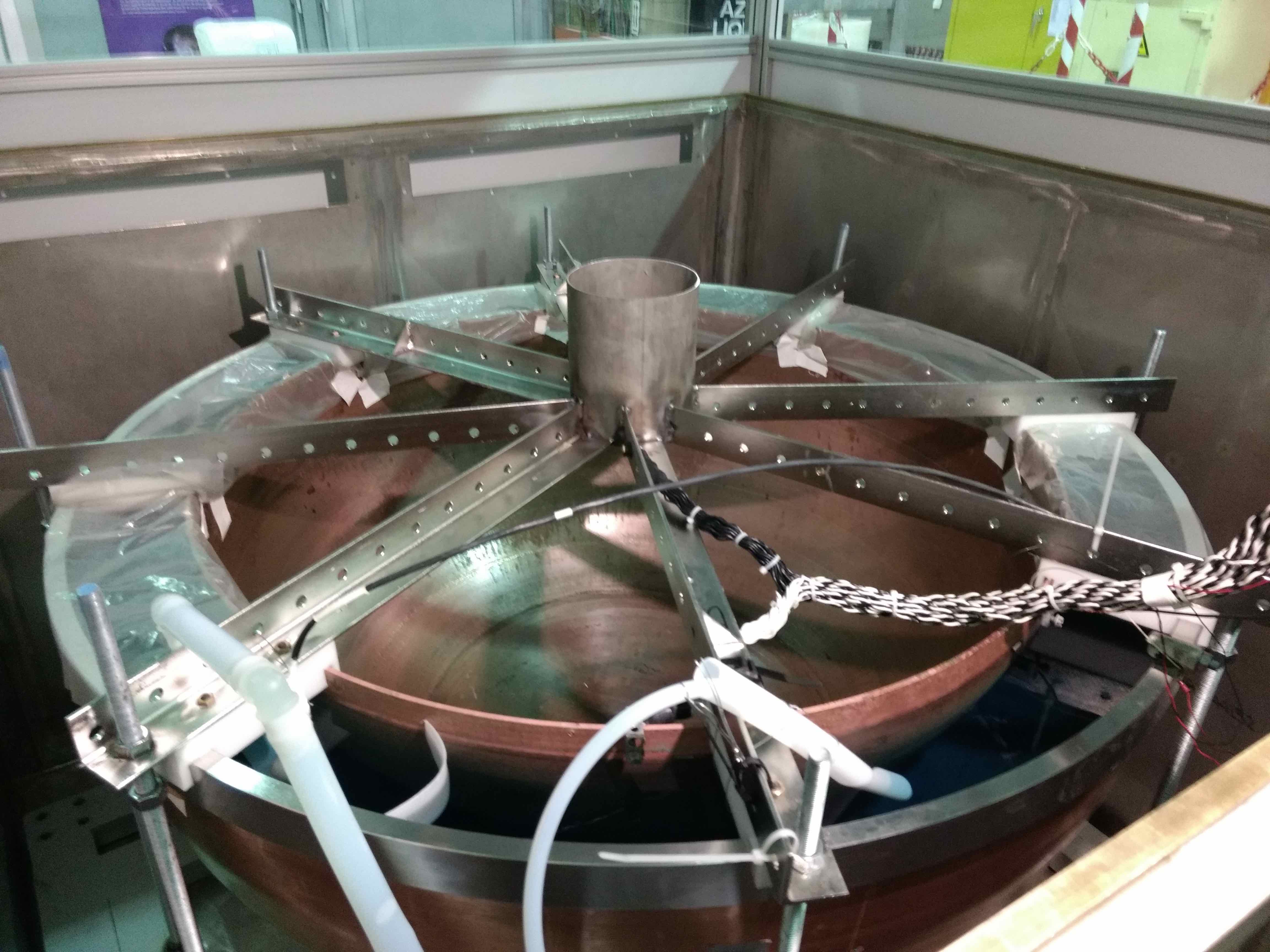}}
\subfigure[\label{subfig:hemisphereElectrolysis}]{\includegraphics[width=1.170\linewidth]{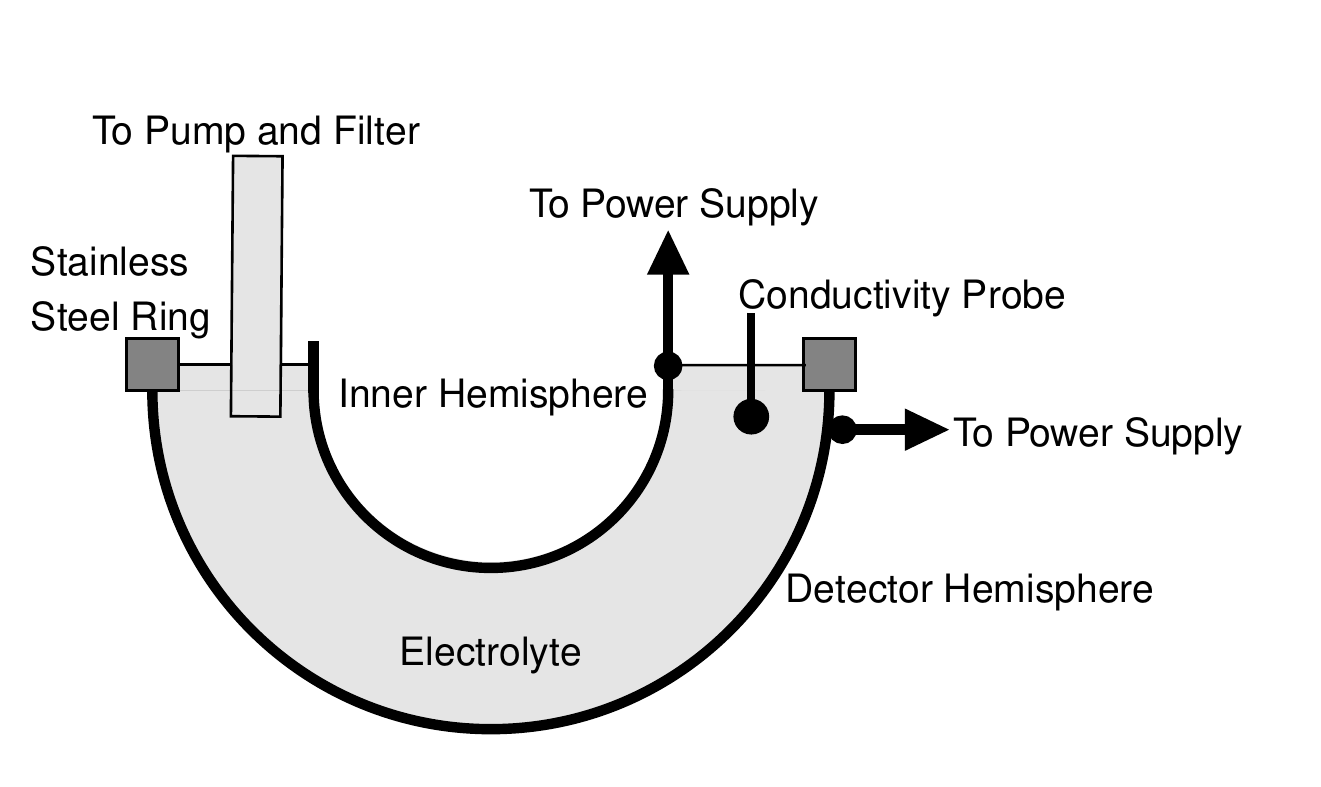}}
 \caption{\subref{subfig:hemi1Installed}\,Electroplating setup showing the detector hemisphere, anode, support structures, and fixtures. \subref{subfig:hemisphereElectrolysis}\,Schematic diagram of the setup.
 \label{fig:hemisphereElectrolysisFigs}}
\end{figure}

A smaller C10100 copper hemisphere was produced to act as the anode for electroplating and was cleaned in the same way as the detector hemispheres. It was suspended inside the detector, separated by an electrolyte comprised of deionised water (18 Mohm), Optima${}^{\text{\textregistered}}$ grade sulphuric acid (Fisher Scientific), and copper sulphate produced by a previous electroplating.
 A pump provided mechanical mixing with a filter removing particulates greater than $1\;\si{\micro\meter}$ in size from the electrolyte. The anode and cathode were connected to a pulse-reverse power supply (Dynatronix, Amery, WI, USA), which could supply up to $80\;\si{\ampere}$.
The whole set-up was contained in a temporary purpose-constructed cleanroom to prevent particulates entering the electrolyte and subsequently providing nucleation sites for nodule-like copper growth~\cite{Overman2012TECHREPmajoranaCopper}. The setup is shown in Fig.~\ref{fig:hemisphereElectrolysisFigs}.

Prior to electroplating, each hemisphere was electropolished to remove a layer of material from the surface. This exposes the underlying crystalline structure and provides an ultra-clean surface prior to deposition. Furthermore, this process enhances the amount of Cu$^{2+}$ in the electrolyte. A higher voltage was used for this process to extract all species from the surface. 
During electropolishing, $(21.2\pm0.1)\;\si{\micro\meter}$ and
$(28.2\pm0.1)\;\si{\micro\meter}$ were removed from the first and
second detector hemispheres, respectively.  This was estimated
from the integrated current and Eq.~\ref{eq:massDeposited}, assuming
uniform polishing.
Following this process, the electrolyte circulated through the filter for several days prior to electroplating to remove particulates released from the copper surface.

\begin{figure}[!h]
\centering
\includegraphics[width=0.85\linewidth]{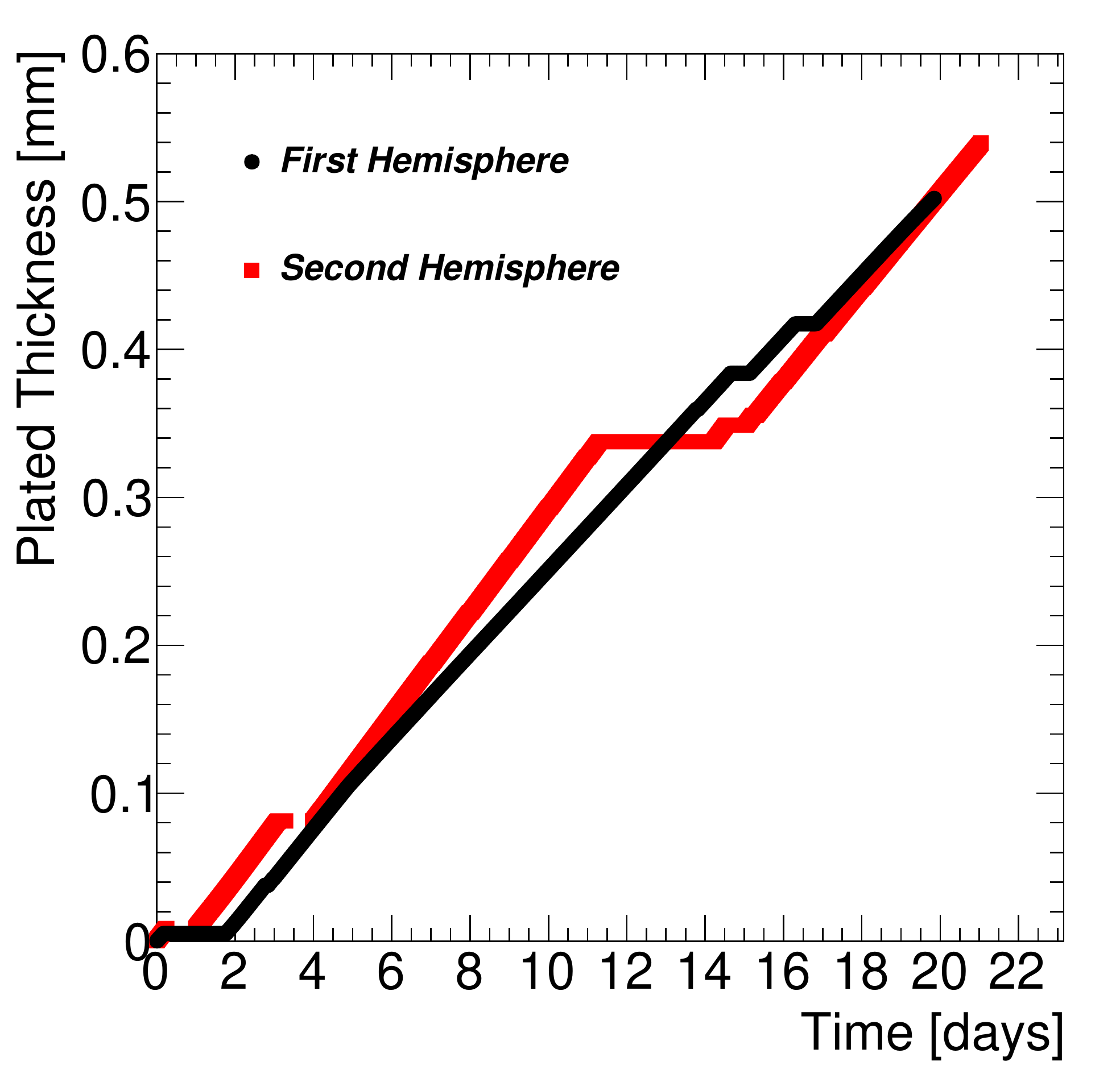}
 \caption{Estimated thickness of the electroplated copper for both detector hemispheres.
 \label{fig:platingRate}}
\end{figure}

\begin{figure}[!h]
\centering
\subfigure[\label{subfig:afterPlating}]{\includegraphics[width=0.49\linewidth]{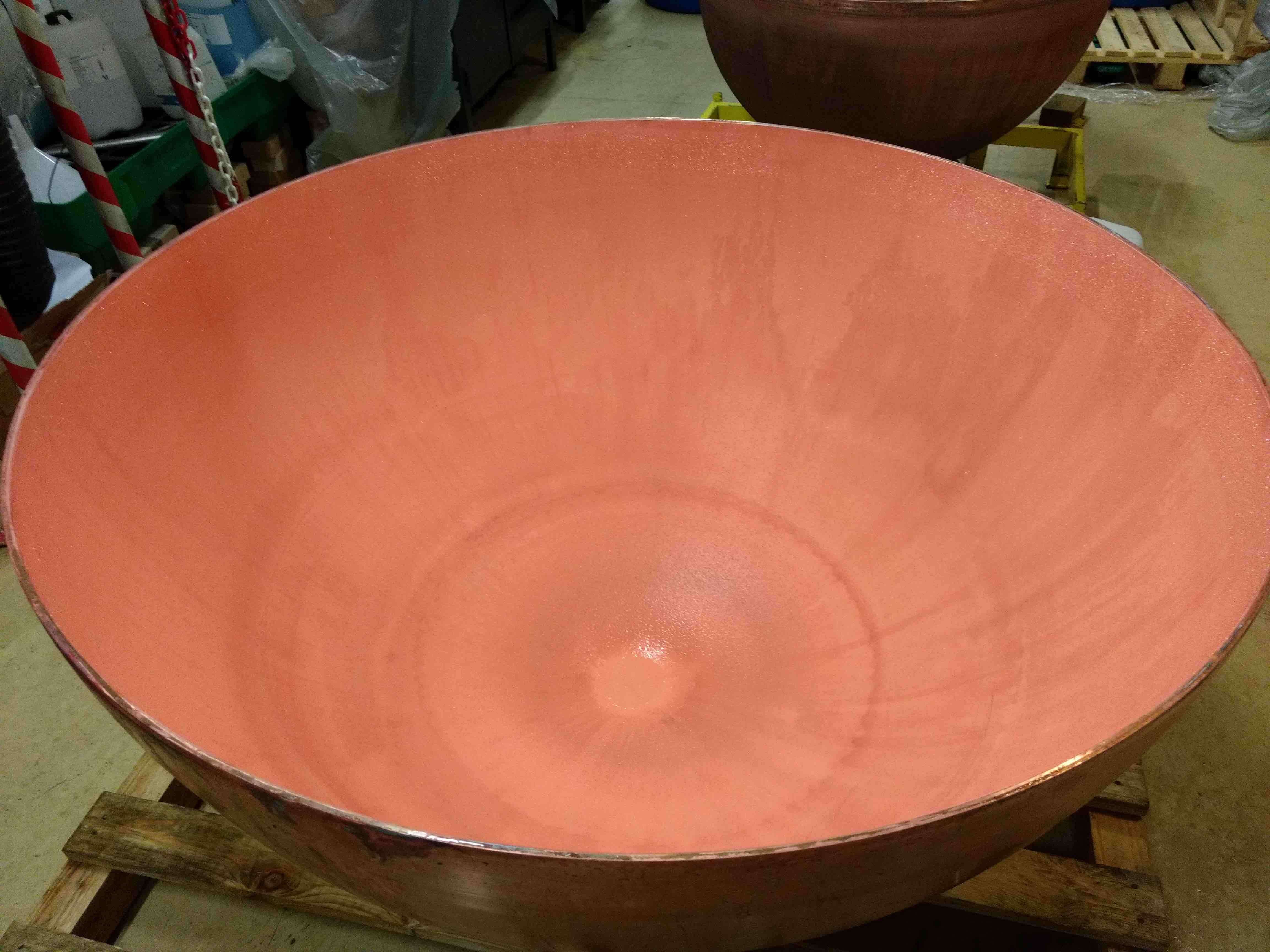}}
\subfigure[\label{subfig:surfaceAfter}]{\includegraphics[width=0.49\linewidth]{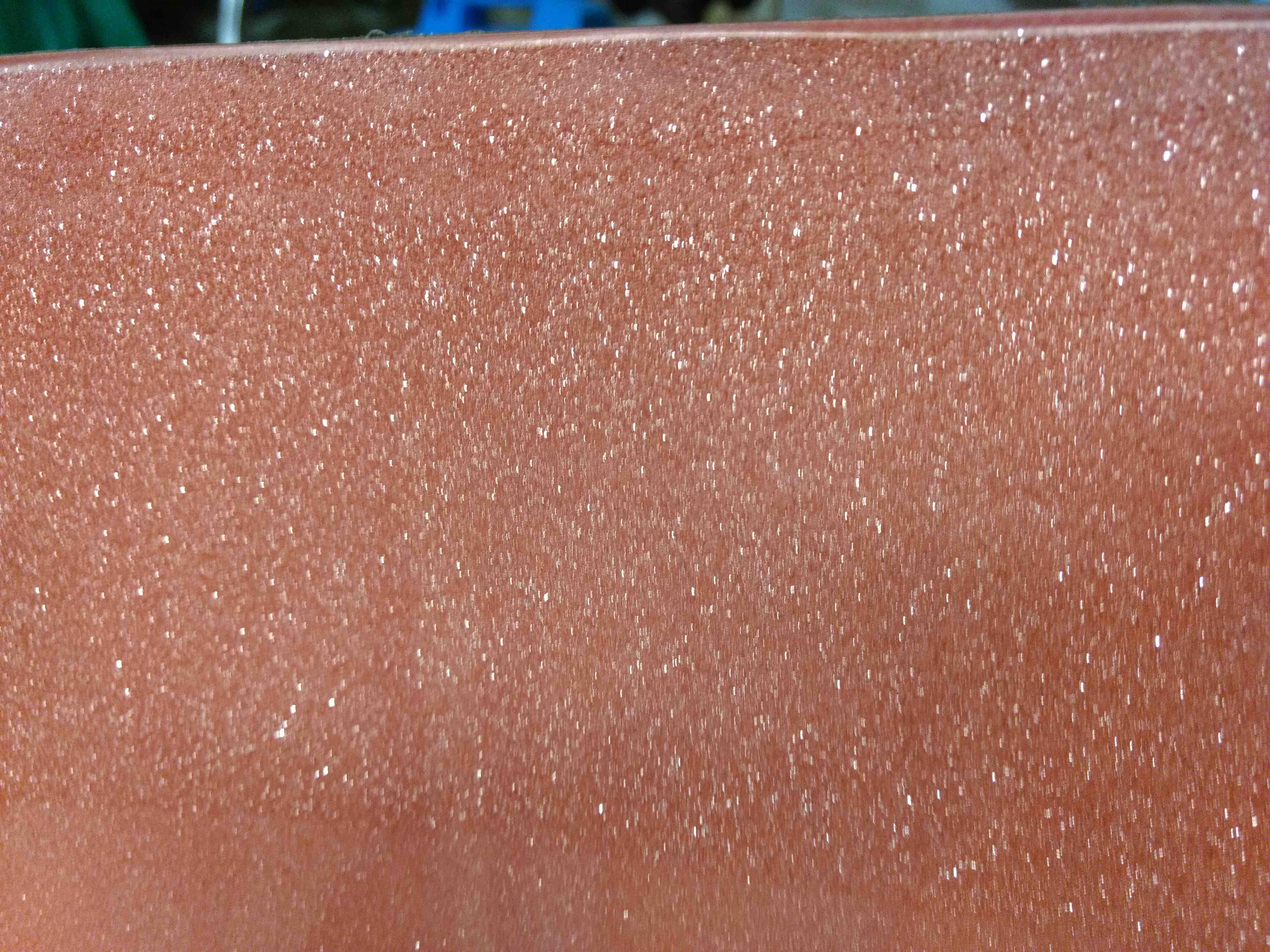}}
 \caption{\subref{subfig:afterPlating} The inner surface of the second hemisphere after electroplating and \subref{subfig:surfaceAfter} a close-up of the surface.
 \label{fig:hemi2Surface}}
\end{figure}

The electroplating procedure used the reverse-pulse plating waveform shown in Fig.~\ref{fig:electrolyticCell}. The current and voltage were monitored throughout, and the conductivity and temperature were recorded using a HACH inductive conductivity sensor.
Electroplating continued for a total of $19.8\;\si{days}$ and $21.0\;\si{days}$ for the first and second hemispheres, respectively. The process took longer than estimated based on the small-scale experiments due to power supply current limitations, plating at slightly lower potential and a slightly lower electrolyte conductivity. This resulted in a slightly more rough surface than obtained in the model. The process was only interrupted for short periods to perform checks or due to power outages. The thickness of the deposited layer, which is shown as a function of time in Fig.~\ref{fig:platingRate}, was estimated from the integrated current assuming a uniform deposition. 
Total copper thicknesses of $(502.1\pm0.2)\;\si{\micro\meter}$ and $(539.5\pm0.2)\;\si{\micro\meter}$ were plated onto the first and second detector hemispheres, respectively. The achieved plating rate corresponds to approximately
$1.3\;\si{\centi\meter\per year}$.
A photo of the finished plating is shown in Fig.~\ref{fig:hemi2Surface}.

After removing the hemispheres from the setup, they were rinsed with deionised water and the surface passivated with a $1\%$ citric-acid solution to prevent surface oxidation~\cite{HOPPE2007486}. Following the welding of the hemispheres together, a final stage of surface etching using an acidified-peroxide solution will be undertaken in order to mitigate the surface contamination caused by contact with the air. This etching technique has been shown to reduce the surface contamination of $^{210}$Pb on electroformed copper to the background level of the XIA UltraLo-1800, which is used in these assays~\cite{Bunker:2020sxw}. This process will be conducted under a nitrogen cover gas to mitigate possible surface recontamination following the etching. Following this stage, the inner detector surface will not be exposed to air again.  

\section{Radioisotope Assay Results}
\label{sec:assayResults}
\begin{figure}[!h]
\centering
\includegraphics[width=0.85\linewidth]{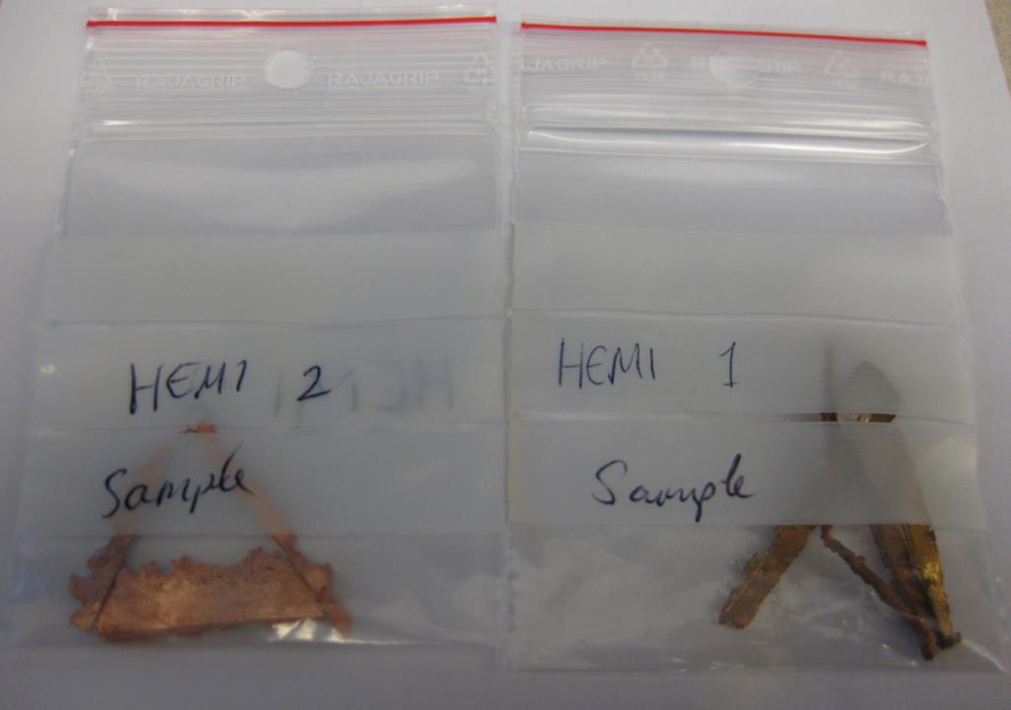}
\caption{Samples of electroplated copper taken from the stainless steel ring shown in Fig.~~\ref{fig:hemisphereElectrolysisFigs}.
\label{fig:copperSample}}
\end{figure}

Samples of the electroplated copper were used to assess its
${}^{238}$U and ${}^{232}$Th concentrations. Samples were taken from
copper plated on the stainless steel ring, shown in
Fig.~\ref{fig:hemisphereElectrolysisFigs}, to avoid damaging the
detector cladding. These samples originate from near the
electrolyte-air interface and from the stainless steel surface;
thus, they represent a worst case scenario with respect to
contamination. The samples collected from each hemisphere are shown in
Fig.~\ref{fig:copperSample}.

The samples were shipped to Pacific Northwest National Laboratory and analysed using ICP-MS following the methods described in Refs.~\cite{LAFERRIERE201593,ARNQUIST2020163761}. 
The results are summarized in Table~\ref{tab:copperPurity}, along with
representative examples of electroformed and commercially sourced
(machined) copper. A substantial improvement over the latter is
observed, with radiopurity levels comparable to previously measured
electroformed copper. The measurement sensitivity for the two
hemispheres is limited
by the mass of the available samples.
Previous samples of electroformed copper have exhibited a 
bulk contamination of $^{210}$Pb lower than the background 
of the XIA UltraLo-1800, and are often used for blank measurements 
or for the construction of the sample tray inside the device~\cite{Bunker:2020sxw,ABE2018157}.  
\begin{table}[!h]
\centering
\caption{ICP-MS results for ${}^{238}$U and ${}^{232}$Th contamination
  in samples of the electroplated copper layer, along with
  representative examples of electroformed and commercially sourced
  copper~\cite{ABGRALL201622}. These are quoted as $68\%$ upper confidence limits, where
  the measurement sensitivity was limited by the available sample
  mass.}
\label{tab:copperPurity}
\vspace{0.5em}
\begin{tabular}{llll}
\hline
       & Weight & ${}^{232}$Th      & ${}^{238}$U \\
Sample &  [g] & [$\si{\microBqperkg}$]       & [$\si{\microBqperkg}$]     \\ \hline
C10100 Cu    &    \multirow{2}{*}{-}     	& \multirow{2}{*}{$8.7\pm1.6$}  & \multirow{2}{*}{$27.9\pm1.9$} \\
  (Machined) & & &  \\
  Cu     &     \multirow{2}{*}{-}     	& \multirow{2}{*}{$<0.119$}  	& \multirow{2}{*}{$<0.099$}   \\ 
  Electroformed    & & &  \\
Hemisphere 1      							 & 0.256	& $<0.58$ 	& $<0.26$ \\
Hemisphere 2      							& 0.614   & $<0.24$  	&$<0.11$  \\ \hline
\end{tabular}
\end{table}

\section{Summary}
The NEWS-G collaboration has utilized recent advances in high-purity
copper electroforming to produce a 
layer of copper on the inner surface of the
$\varnothing 140\;\si{\centi\meter}$ detector.  This layer will act as
a shield to mitigate background from $^{210}$Pb in the bulk of the
detector's commercially sourced C10100 copper. This is the largest surface
to be plated with ultra-radiopure copper in an underground
laboratory. This operation has demonstrated the feasibility of plating
onto the surface of a large hemisphere. The radiopurity of the plated
copper was assessed using ICP-MS and found to be comparable to other
electroformed copper. A copper deposition rate of approximately
$1.3\;\si{\centi\meter\per year}$ was achieved, which is promising for
fabrication of a fully electroformed copper sphere in the future.

\section*{Acknowledgments}
A portion of this work was funded by PNNL Laboratory Directed Research
and Development funds under the Nuclear Physics, Particle Physics,
Astrophysics, and Cosmology Initiative.  The Pacific Northwest
National Laboratory is a multi-program national laboratory operated
for the U.S. Department of Energy (DOE) by Battelle Memorial Institute
under contract number DE-AC05-76RL01830.
This project has received funding from the European Union's Horizon
2020 research and innovation programme under the Marie
Sk\l{}odowska-Curie grant agreement DarkSphere (grant agreement
No~841261). Support has been received from the Royal Society
International Exchanges Scheme.
This research was undertaken, in part, thanks to funding from the
Canada Excellence Research Chairs Program, the Canada Foundation for Innovation, the Arthur B. McDonald
Canadian Astroparticle Physics Research Institute, and the French
National Research Agency (ANR-15-CE31-0008).
The authors would like to thank the XMASS collaboration for the use of
their XIA detector through the NEWS-G/XMASS collaborative agreement.
\bibliography{mybib} 
\end{document}